\documentstyle[prl,aps,epsf,multicol]{revtex}
\begin{document}

\raggedcolumns
\begin{multicols}{2}
\narrowtext

{\bf Reply to a Comment on ``Nonequilibrium Electron 
Distribution in Presence of Kondo Impurities'' 
(cond-mat/0105026)} 


In a recent paper \cite{GeorgErelaxPRB01} 
we have studied the energy relaxation of electrons in 
voltage biased mesoscopic wires in presence of magnetic 
impurities. The t-matrix approach by Kaminski and Glazman (KG) 
\cite{GlazmanERelaxPRL01} was extended beyond the poor 
man's scaling regime and shown to lead to results 
in quantitative agreement 
with experimental data by Pothier {\it et al.}
\cite{Pothier} and Pierre {\it et al.} \cite{Pierre}. 

Kroha and Zawadowski (KZ) argue in a Comment 
\cite{KrohaComm01} that the decomposition of the 
two particle t-matrix into two single particle 
t-matrices employed by KG 
and us is not adequate. As shown in \cite{GeorgErelaxPRB01} 
such a decomposition arises for the 
leading infrared divergent terms of the collision kernel. 
These terms display a $1/ \omega^2$ behavior, 
where $\omega$ is the energy exchanged between the two
 electrons, which is responsible for 
the experimentally observed scaling of the nonequilibrium 
electron distribution function. 
As already noted by KG, the $1/ \omega^2$ behavior is cut off 
at low energies by the decoherence rate $1/ \tau_s$ 
of the impurity spin. On the other hand, in the collision 
integral the collision kernel is multiplied by a product 
of four distribution functions which 
vanishes for low energies 
so that the collision integral is in fact well behaved 
even in the absence of an infrared cutoff. As a consequence, 
the energy range $\omega < 1/ \tau_s$ where our approach may
overestimate the collision kernel gives only a 
negligible contribution to the collision integral, and 
the spin decoherence rate leads merely to secondary effects 
that do  not influence the scaling behavior.
Various tests with cutoffs based on estimates by KG have 
confirmed this conclusion. 

While this was already 
explicitly mentioned in \cite{GeorgErelaxPRB01} KZ 
now argue that for typical experimental parameters an 
infrared regularization is required in about 10 to 30$\%$ 
of the relevant energy range and, therefore, summing the
leading divergent terms in $1/\omega$ is not sufficient.  
However, KZ seem to have missed the selfconsistency of 
our approach. 
As can be seen from Fig.~3 in \cite{GeorgErelaxPRB01} 
the selfconsistently determined single particle spin-flip 
t-matrix $\tau (\varepsilon)$ 
does not vary much in magnitude in the relevant energy 
range since it depends on the smeared nonequilibrium electron 
distribution function and not on the distribution of 
noninteracting electrons with two sharp Fermi edges. 
For sufficiently smeared distribution 
functions the argument by KZ is not correct since
nonleading divergent terms are suppressed.
Even in presence of a cutoff the  
factorized graphs give the leading contribution
and our approach is indeed adequate to describe the 
experiments in \cite{Pothier,Pierre}. 
In fact, our fit of the gold data 
\cite{Pierre} has no adjustable parameters since the 
impurity concentration was extracted from the temperature 
dependence of the resistivity. 

In \cite{GeorgErelaxPRB01} we point out that our 
result ``is insensitive to the Kondo 
temperature $T_K$ as long as $T \ll T_K$''. 
$T_K$ can in fact  be varied within a 
physically reasonable regime determined by the
experimental conditions. KZ correctly point 
out that our approach does not 
describe the crossover to the Fermi liquid fixpoint behavior. 
This issue was already discussed by KG in some detail and 
is of no importance in the 
parameter regime of the experiments in \cite{Pothier,Pierre}.  

In a last point KZ write ``G\"oppert and Grabert claim 
that in the slave boson (SB) method employed in 
\cite{KrohaASSP00,KrohaNEQPR01}  algebraic behavior of 
$K( \omega,\varepsilon,\varepsilon^{\prime})$ can only be 
obtained in infinite order perturbation theory (PT)''. 
This claim has never been made by us. 
Ref.~\cite{KrohaNEQPR01} is even not cited in
\cite{GeorgErelaxPRB01} and hardly could be mentioned, since 
it has appeared in cond-mat only  2 months later than 
\cite{GeorgErelaxPRB01}. In the published 
version of \cite{GeorgErelaxPRB01}, we have added a reference to 
\cite{KrohaASSP00} in a sentence referring to the
$1/\omega^2$-behavior and
``the NCA techniques used in \cite{KrohaASSP00} 
where algebraic behavior only arises from a summation of an 
infinite series of logarithmic corrections.''  
Hence, it was not claimed that the SB approach {\it a priori} 
cannot reproduce the algebraic $1/ \omega^2$ behavior 
in PT but rather that Kroha's specific 
calculations do not yield algebraic $1/ \omega^2$ terms. 
This is in complete accordance with 
Kroha's own statements in \cite{KrohaASSP00} 
where he writes about the $1/ \omega^2$ term 
responsible for scaling: ``It must, therefore, be generated 
by an infinite resummation of logarithmic terms obtained 
in perturbation theory due to the presence of a Fermi edge.''  
Basically, we have just restated  
Kroha's own words. Of course, our results can also be 
reproduced by other methods. This work was supported
by grants from the DFG and the DAAD.

We wish to thank B.~Altshuler, F.~Pierre, and H.\ Pothier 
for valuable discussions.

\vspace*{0.5cm}

\noindent
Georg G\"oppert \hfill Hermann Grabert \\
Princeton University \hfill Fakult\"at f\"ur Physik\\
Jadwin Hall, PO Box 708 \hfill Albert--Ludwigs--Universit{\"a}t\\ 
Princeton, NJ 08544 \hfill 79104 Freiburg, Germany\\

%
%

%
\end{multicols}
\end{document}